\documentclass[aps,pre,twocolumn,superscriptaddress]{revtex4-1}
\usepackage[english]{babel}
\usepackage{graphicx}
\usepackage{amsmath}
\usepackage{amsfonts}
\usepackage[usenames,dvipsnames]{color}

\usepackage{hyperref}
\usepackage{rotating}
\usepackage{booktabs} 
\usepackage{slashed}
\usepackage{transparent}
\usepackage{color}
\usepackage{soul}
\usepackage{tikz}

\usetikzlibrary{external}
\usetikzlibrary{arrows}

\usetikzlibrary{patterns}
\usetikzlibrary{shapes,snakes}

\newcommand{\dd}{\; \mathrm{d}}

\DeclareMathAlphabet{\mathbb}{U}{bbold}{m}{n}

\newcommand{\bs}[1]{\boldsymbol{#1}}

\usepackage[normalem]{ulem}

\usepackage{hyperref}
\usepackage{nameref}

\newcommand{\probability}{\mathbb{P}}
\newcommand{\expectation}{\mathbb{E}}

\begin{document}

\title{Polydispersity in Percolation}

\author{Fabian Coupette}
\email{f.coupette@leeds.ac.uk}
\affiliation{School of Mathematics, University of Leeds, Leeds LS2 9JT, United Kingdom}
\affiliation{Institute of Physics, University of Freiburg, Hermann-Herder-Stra{\ss}e 3, 79104 Freiburg, Germany}
\author{Tanja Schilling}
\email{tanja.schilling@physik.uni-freiburg.de}
\affiliation{Institute of Physics, University of Freiburg, Hermann-Herder-Stra{\ss}e 3, 79104 Freiburg, Germany}

\date{\today}

\begin{abstract}

{
Every realistic instance of a percolation problem is faced with some degree of polydispersity, e.g., the pore-size distribution of an inhomogeneous medium, the size distribution of filler particles in composite materials, or the vertex degree of agents in a social network. Studies on different classes of systems have independently found very similar conceptual results for the percolation problem, i.e., that the percolation threshold is insensitive to the particular distribution controlling the polydispersity. Rather, the percolation threshold depends only on the first few moments of the distribution. In this article, we explain this frequently observed pattern using branching processes. The key observation is that a reasonable degree of polydispersity effectively does not alter the structure of the network that forms at the percolation threshold. As a consequence, the critical parameters of the monodisperse system can be analytically continued to account for polydispersity.
}

\end{abstract}

\maketitle

In this article, we discuss the interplay of particle size polydispersity and connectivity percolation thresholds. Across a variety of different systems, it has been observed that details of size distributions do not affect the percolation transition: The emergence of a giant component appears to depend only on low-order moments of the corresponding distribution. In particular, this has been demonstrated for slender rod-like particles \cite{otten2009continuum,chatterjee2010connectedness,otten2011connectivity,mutiso2012simulations,nigro2013quasiuniversal,meyer2015percolation,majidian2017role}. These are frequently studied as a model system for carbon nanotubes, which play an important role in the fabrication of conductive and piezo-resistive elastomer composites. The same application has recently also inspired the analysis of ensembles of fractal aggregates as a model for carbon-black-filled composite materials \cite{coupette2021percolation,chatterjee2022geometric}. Even though the length of carbon nanotubes can be controlled well in the fabrication process, any realistic composite will be subject to some level of polydispersity. 
Thus, it is crucial to understand the implications of polydispersity in order to ultimately optimize the product. Apart from immediate industrial interest, polydispersity has been studied in a variety of different theoretical model systems, such as two-dimensional sticks or disks in the plane  \cite{quintanilla2001measurement,quintanilla2007asymmetry,chatterjee2014percolation,meeks2017percolation,tarasevich2018percolation,bissett2023percolation}. Furthermore, complex networks with inhomogeneous vertex degrees, such as modified Bethe lattices or correlated networks, have been examined as the discrete analogue to particle polydispersity \cite{goltsev2008percolation,widder2019high,li2021percolation}. 
All of the aforementioned systems have been analyzed with system-specific methodology, such as connectedness percolation theory, excluded-volume theory, lattice mappings, message passing, computer simulation, and conductivity measurements. A recurring pattern in the collective findings is that the critical parameters in most instances can adequately be represented as a function of only mean $\mu$ and variance $\sigma^2$ of the underlying size distribution.
In this work, we elucidate the common reasons for this seemingly universal trend.

Our central message is that the percolation threshold for most standard size distributions can be readily inferred from a monodisperse reference system,  which we construct in the following. 
The key observation is that small particles have negligible impact on the formation of a percolating network unless they vastly outnumber the larger particles. The distributions that are commonly studied theoretically or observed experimentally are rarely skewed enough for the network structure of the polydisperse system to meaningfully deviate from the network structure of a monodisperse system. This allows us to infer the percolation threshold from the first moments of the underlying size distribution.

In Section~\ref{sec:1}, we recall percolation on treelike networks with varying degree distributions in terms of branching processes. We proceed by illustrating how percolation on a planar lattice can be cast as a branching process and analyzed accordingly in Section~\ref{sec:2}. We then examine the continuum problem of bidisperse circles in section \ref{sec:3}, demonstrating the type of distribution that violates the common pattern. The affected distributions require a higher-order treatment that is comparatively tedious, as illustrated by the second-order treatment of the circle system. However, such treatment is rarely necessary, which we demonstrate by surveying standard model systems that vary in particle shape, dimension, and interaction in Section~\ref{sec:4}.

\section{Arboreal Polydispersity}\label{sec:1}

\begin{figure}[h]
	\centering
	\includegraphics[width=0.49 \textwidth]{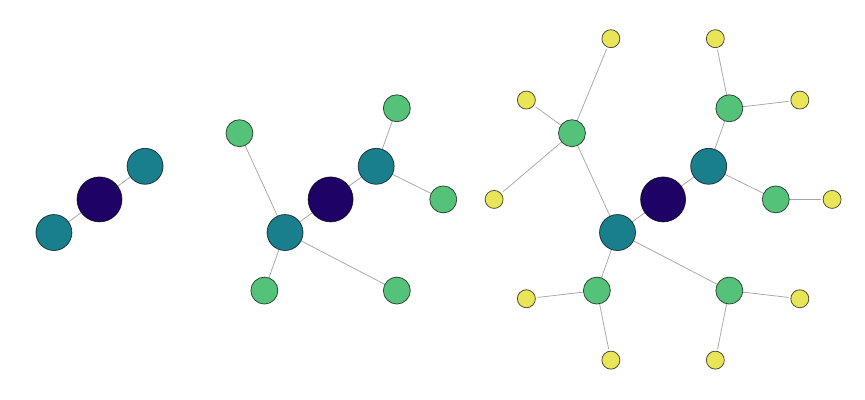}
	\caption{Iterative generation of a treelike network with a prescribed vertex degree distribution from a branching process. A branching process is initiated at a root vertex (dark blue) -- its degree is drawn randomly from the vertex degree distribution $\probability(z)$.  For each neighbor of the origin we add neighbors drawn randomly from $\probability(z-1)$ as the origin already contributes to the degree and continue in this fashion indefinitely. Vertex colors and sizes highlight the iteration of the branching process a vertex is generated.    }
	\label{fig:intro}
\end{figure}    

In order to provide a reference for more complicated scenarios, we begin by considering bond percolation on a treelike network. Bonds are ``open'' or ``closed'' with fixed probabilities, $p$ and $1-p$. Percolation generally corresponds to the formation of a ``giant'' component of connected open edges, but we need to be more precise about what this means to avoid confusion arising from different definitions across network science and mathematical literature. 

A mathematically rigorous definition of the percolation transition requires an infinite system. 
Thus, given an \emph{infinite} system, we define the percolation probability $\Theta_v$ as the probability that a specific vertex $v$ (``origin'') is part of an infinite component. With that, the percolation threshold can be defined as
\begin{align}
p_c = \sup \{p \, | \, \Theta_v(p) = 0 \} \; ,
\label{eq:def}
\end{align}
i.e., the largest parameter such that the cluster around the origin remains almost surely finite.
As long as there is a finite probability to link any two vertices in the network through a series of open edges, the percolation threshold does not depend on the choice of origin, while the percolation probability $\Theta_v$ for $p > p_c$ generally varies with $v$. 
Any infinite simply connected graph can serve as the basis of a percolation model with a well-defined  percolation threshold. 

Note that the networks commonly studied in explicit applications are often \emph{finite}, such that there will never be a percolating cluster according to our definition. 
In this case, percolation instead depends on how the size of the largest cluster scales with the size of the network. If the size of the largest cluster diverges with the system size, the original system is deemed percolating. However, this requires a notion of ``upscaling'' the original graph. 
Contact networks inferred from a continuum distribution of particles in space are upscaled simply by extending the underlying space while keeping the intensive physical parameters, such as the particle density, fixed. Yet, in other cases the upscaling protocol may not be unique and needs to be specified explicitly. 


First, we cast the percolation problem as a branching process. (We have described this method in detail in Ref.~\cite{coupette2023universal}; here, we just recall the basic idea.) We denominate an arbitrary vertex as the ``origin'' and call a vertex ``open'' if it is connected to the origin through a path of open edges. Due to the absence of loops, there is exactly one self-avoiding path between any two vertices in the network. The number of edges comprising this path can serve as a measure of distance on the graph, partitioning the network into ``concentric layers'' $V_n$ of vertices at the same distance to the origin. 
``Polydispersity'' for such a discrete system translates to a heterogeneous distribution of coordination numbers $z$ across the vertices of the network. Assume we draw vertex degrees independently from a common distribution and consider percolation on the ensemble of all infinite treelike networks constructable in this way. 
Following the terminology we developed previously \cite{coupette2022exactly}, we can start at a randomly chosen origin and describe the number of open vertices in the layers $V_n$ as the branching process $(X_{n})_{n\in \mathbb{N}}$ defined through
\begin{align}
X_{n+1} = \sum_{i=1}^{X_n} \xi_i^n \;,
\label{eq:branching}
\end{align}
with $\xi_i^n$ denoting independent random variables that capture the number of open vertices in $V_{n+1}$ induced by the $i^{\mathrm{th}}$ open vertex in $V_n$. For example, the origin in Fig.~\ref{fig:intro} has two neighbors, such that the number of open neighbors is a binomially distributed random variable, $\xi^0_1 \sim B(2,p)$. More generally, for all $n>0$, we have $\xi^n_i \sim B(z^n_i-1,p)$,  with $z^n_i$ denoting the coordination number of the corresponding vertex. Knowing the vertex-degree distribution $\probability(z)$, we can compactly write
\begin{align}
   \xi^{n>0}_i \equiv \zeta \sim \sum_z \probability(z) B(z-1,p) \, ,
\end{align}
as the r.h.s does not depend on $i$ or $n$ anymore. Generally we define $\zeta$ as the random variable describing branching of a vertex sitting at the surface of the growing cluster around the origin but infinitely far away. Explicitly $\zeta$ describes the number of open vertices in $V_{n+1}$ induced by any single open vertex in $V_n$ for $n \rightarrow \infty$.
We further define $f_\zeta$ as the corresponding probability generating function. 
Conveniently, for a branching process of the form in eq.~\eqref{eq:branching}, the probability generating function of $X_{n+1}$, i.e., $f_{X_{n+1}}$, is given by
\begin{align}
	f_{X_{n+1}} = f_{X_{n}} \circ f_\zeta \;.
\end{align}
We can cast this recurrence in a closed form
\begin{align}
	f_{X_{n+1}} = f_{X_{1}} \circ f^{(n)}_\zeta \;,
\end{align}
featuring the $n$-fold concatenation of $f_\zeta$. As $f_\zeta$ is identical for all layers beyond the first one (the origin itself has a different distribution of branching targets), the survival or termination of the branching process depends exclusively on $f_\zeta$. 
More precisely, the termination probability 
\begin{align}
	Q = \lim_{n \rightarrow \infty} f^n_\zeta(0)
\end{align}
 has to be a fixed point of $f_\zeta$. As normalization requires $f_\zeta(1) = 1$, $Q = 1$ (certain doom) is always a fixed point. Yet, this fixed point becomes unstable as a second fixed point emerges when the fixed point at 1 becomes tangential, i.e., $f'_\zeta(1) = 1$, which corresponds to the simple criticality criterion
 \begin{align}
 	\expectation[\zeta]  = 1 \; .
    \label{eq:crit1}
 \end{align} 
If the branching process terminates, the cluster containing the origin is finite. Conversely, if the branching process does not almost surely terminate, there is a non-vanishing probability that the origin is part of an infinite component, which means the system is percolating. 
Simply put, eq.~(\ref{eq:crit1}) says the system is critical if any open vertex on average generates another open vertex on the next layer. 
The left hand side encodes the average number of open branching targets of an open vertex far away from the origin which in our simple example is simply $\langle z \rangle - 1$ with $\langle z \rangle = \sum_z z \probability(z)$.
Accordingly, the percolation threshold is given by
\begin{align}
	p_c = \frac{1}{\langle z \rangle - 1}\;, 
    \label{eq:pc1}
\end{align} 
which implies that all distributions with the same mean yield the same percolation threshold.
The specific distribution of vertex degrees is entirely irrelevant for percolation as criticality depends exclusively on the first moment of the distribution.
In that sense, polydispersity has no effect on the percolation threshold.

Notably, eq.~(\ref{eq:pc1})  differs from the well-established Molloy-Reed criterion for random graphs with uncorrelated vertex degrees \cite{molloy1995critical,hamilton2014tight,karrer2014percolation}
\begin{align}
	p_c = \frac{\langle z \rangle}{\langle z^2 \rangle - \langle z \rangle}\; .
    \label{eq:pc2}
\end{align} 
Both results are correct with respect to their corresponding graph ensembles but it is important to realize that eq.~(\ref{eq:pc2}) is a special case of eq.~(\ref{eq:crit1}) and indeed \emph{not} a general solution to percolation on treelike networks. Eq.~(\ref{eq:pc2}) applies for the configuration model, i.e., random graphs generated from independently drawn vertex degrees (every vertex has a preset number of outgoing edges). Randomly matching pairs of outgoing edges creates an ensemble of locally treelike networks. However, the matching process skews the degree distribution of nearest-neighbors because there are simply more matching pairs connecting to a vertex of high degree compared to a vertex with few outgoing edges. Thus, the degree distribution of a random neighbor of the origin is not $\probability(z)$ but rather the \textit{excess degree distribution}
$\probability_{\mathrm{ex}}(z) = \frac{z}{\langle z \rangle} \probability(z)$. The distribution of branching targets for all vertices except the origin and hence also  $\zeta$ is controlled by $\probability_{\mathrm{ex}}$. Accordingly, eq.~(\ref{eq:crit1}) yields
\begin{align}
    \expectation[\zeta] &=& & p \sum_z (z-1) \probability_{\mathrm{ex}}(z) = p \sum_z (z-1) \frac{z}{\langle z \rangle}\probability(z) \nonumber \\ &=& &p \frac{\langle z^2\rangle -\langle z \rangle}{\langle z \rangle} = 1
\end{align}

If we instead generate a network through a branching process, degree distributions of origin and neighbor coincide.
As both models describe an ensemble of in some sense uncorrelated (locally) treelike networks yielding different results it is important to be precise in defining the ensemble.
Notice that if we take the average of $z$ in eq.~\eqref{eq:pc1} with respect to the excess degree distribution, we retrieve
\begin{align}
p_c = \frac{1}{\langle z \rangle_{\mathrm{ex}} - 1} = \frac{1}{\langle z \frac{z}{\langle z \rangle} \rangle -1} = \frac{1}{\frac{\langle z^2 \rangle - \langle z \rangle}{\langle z \rangle}} = \eqref{eq:pc2}\;, 
\end{align}
the standard result. Thus, eq.~(\ref{eq:pc1}) is generally true but the specific model may modify the definition of the average $\langle \cdot \rangle$.
This subtlety is lost in the continuum, as we will see below, but it should be noted that the definition in eq.~\eqref{eq:def} and the corresponding percolation threshold in eq.~\eqref{eq:pc1} also work for discrete systems, despite the superficial conflict with the conventional wisdom. Eq.~\eqref{eq:pc2} is a special case of eq.~\eqref{eq:pc1}, accounting for a biased degree distribution perceived on a self-avoiding random walk across the network. 
Irrespective of the network generation protocol, vertex degrees may be correlated explicitly. In the branching framework, correlations change the properties of the indefinitely reiterated probability generating function $f_\zeta$.  
We define $\chi := z - 1$ as the branching ratio, i.e., the non-backtracking coordination number~\cite{goltsev2008percolation}. In case of nearest-neighbor correlations, the degree sequence encountered on a self-avoiding random walk on a treelike network is generated by a Markov process driven by the branching matrix 
\begin{align}
\Lambda := (l P_{kl})_{kl} = (l \probability(\chi_2 = l | \chi_1 = k))_{kl}.
\end{align}
In order to capture the asymptotic growth of the network, we need to evaluate arbitrary powers of $\Lambda$.  
The largest eigenvalue $\lambda_1$ of this matrix characterizes how many neighbors a vertex in $V_n$ will have on average in $V_{n+1}$ for $n \rightarrow \infty$. 
The corresponding properly normalized eigenvector encodes the probability distribution of branching numbers in the steady state of the Markov process.  
For example, on a tree with alternating branching numbers $\chi_1$ and $\chi_2$, a generation after two successive branching steps has expanded by a factor of $\chi_1 \chi_2$. 
This corresponds to a growth per generation of $\sqrt{\chi_1 \chi_2}$ (which is also the largest eigenvector of $\Lambda$) and thus $p_c = \frac{1}{\sqrt{\chi_1 \chi_2}}$. In contrast to that, a homogeneous distribution of branching numbers yields the threshold $p_c = \frac{2}{(\chi_1 + \chi_2)}$.
Any treelike percolation problem simplifies to determining the asymptotic growth rate $R$ as a function of the control parameters, and criticality occurs at $R = 1$.
Correlations of finite range can be integrated out and correlated polydispersity may have an impact on the percolation threshold. However, the plain Bethe lattice with constant coordination number will always maintain the lowest percolation threshold among all treelike lattices with the same mean coordination number. This boils down to appreciating that the product $\prod_i a_i$ is maximized by a homogeneous distribution of $a_i$'s under the constraint that $\sum_i a_i$ is constant. Thus, a treelike lattice grows most efficiently in a homogeneous manner. This is an important observation because it persists when introducing loops.

\section{An Instructive Example}
\label{sec:2}

We now introduce polydispersity to the square lattice $\mathbb{Z}^2$ while keeping the bond percolation problem exactly solvable. Each edge on the square lattice becomes a degree of freedom which is either open with probability $p$ or closed with probability $1-p$. The rigorous calculation of the exact percolation threshold $p=\frac{1}{2}$ on the basis of lattice duality is one of the most celebrated results in the mathematical research on percolation \cite{kesten1980critical}.

\begin{figure}[h]
	\centering
	\includegraphics[width=0.33 \textwidth]{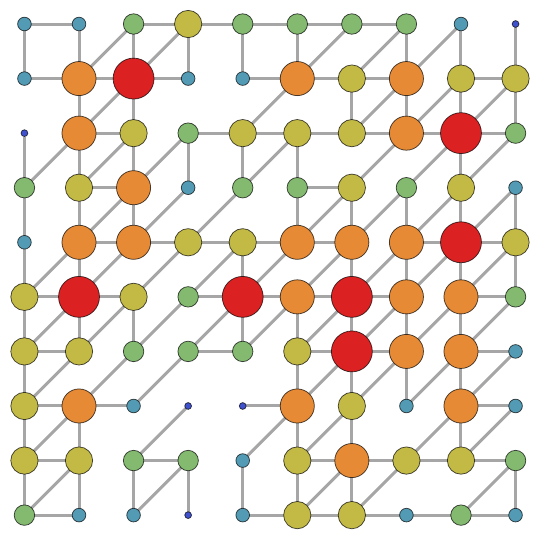}
	\caption{Square lattice with a third of its edges exchanged for diagonals, resulting in an inhomogeneous degree distribution indicated by size and color of each vertex.}
	\label{fig:psql}
\end{figure}

 This result has been extended to a broader variety of lattices including all instances of the so-called triangular hypergraph (see Fig. \ref{fig:psql}), e.g., the triangular lattice $\mathbb{T}^2$ \cite{fisher1961some,sykes1964exact,ziff2006exact}. The triangular lattice naturally includes the square lattice as a subgraph by taking only two of the sides of any upward-pointing triangle (always the same two). In that sense, the square lattice can be considered a triangular lattice with correlated defects.  

The square lattice has a uniform vertex degree of 4. We can introduce polydispersity by randomly replacing an edge of $\mathbb{Z}^2$ by one of $\mathbb{T}^2 \setminus \mathbb{Z}^2$. In that way, the average coordination number remains unaltered while the model also remains exactly solvable: The problem maps to percolation on a triangular lattice with individual probabilities for each of the three bonds comprising a triangle, a problem that still allows for an exact solution. Thus, for this model, we can study the correlation between polydispersity, i.e.,~vertex-degree distribution, and percolation threshold in exact terms---a rare opportunity for non-trivial systems.

\begin{figure}[h]
	\centering
	\includegraphics[width=0.49 \textwidth]{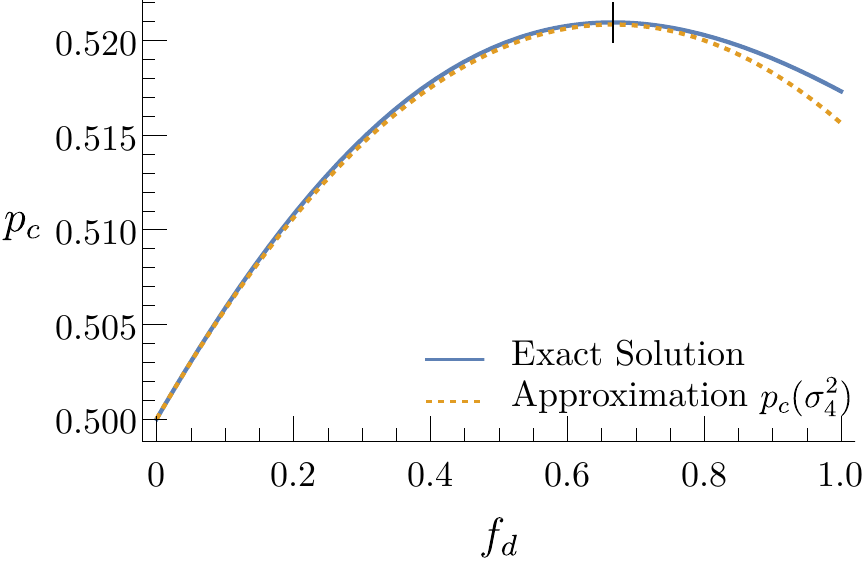}
	\caption{Percolation threshold of a polydisperse square lattice with mean coordination number $\langle z \rangle = 4$. The dashed line displays the first-order expansion of $p_c$ in $f_d$ around the square lattice ($f_d = 0$).  }
	\label{fig:example}
\end{figure} 

Define $f_s$ as the probability that an edge of the original square lattice exists as part of the rewired lattice, and $f_d$ as the corresponding probability for added diagonal junctions. The probability that an edge is open remains $p$ for all edges. Therefore, the probability that an edge of the original square lattice exists and is open is $p_s =  p f_s$ and analogously for the diagonals $p_d =  p f_d$. It is important to notice that for bond percolation, closed edges are equivalent to non-existent edges. Thus, we may simply consider a complete triangular lattice with openness probabilities $p_s$ and $p_d$ for the respective edges. As shown in Ref.~\cite{coupette2022exactly}, this lattice percolates if 
\begin{align}
	2 p_s + p_d - p_d p_s^2 = 1 \; ,
\end{align}
which implies
\begin{align}
	p(f_d+f_s(2-f_s f_d p^2) ) = 1 \; .
	\label{eq:fsnps}
\end{align}
The mean vertex degree of the lattice blueprint is given by
\begin{align}
	\langle z \rangle = 2(2 f_s + f_d) \; .
\end{align}
Replacing $f_s$ by the average vertex degree in eq.~\eqref{eq:fsnps} yields 

\begin{align}
  R(p, \langle z \rangle, f_d) := \frac{p \langle z \rangle}{2} -\frac{1}{4}\left(\frac{\langle z \rangle}{2}-f_d\right)^2 f_d p^3 = 1 \; ,
  \label{eq:example}
\end{align}
with $\langle z \rangle$ and $f_d$ such that $0 \leq f_s, f_d \leq 1$. 
This result allows for a couple of important observations. 
First, treating $f_d$ as a small perturbation, the ``mean-field'' percolation threshold depends exclusively on the mean vertex degree. 
Furthermore, since the second term on the l.h.s. is strictly negative for $f_d \in (0,1]$, the percolation threshold for a fixed mean degree is minimal for globally homogeneous vertex degree, i.e., $f_d = 0$. We can compute the derivative 
\begin{align}
	\frac{d p_c}{d f_d} = -\frac{\partial R}{\partial f_d} \left( \frac{\partial R }{\partial p} \right)^{-1}
	\label{eq:imp}
\end{align}
of the implicit function $p_c(f_d,\langle z \rangle)$ at a solution of eq.~\eqref{eq:example}, for example, the monodisperse system characterized by the triplet $(\frac{1}{2},4,0)$. The result can be expanded in $f_d$ to estimate the response of the percolation threshold to small parameter adjustments. In the same way, we may expand the variance of the degree distribution
\begin{align}
	\sigma_z^2 = \langle z^2 \rangle -\langle z \rangle^2 =\langle z \rangle(1+ f_d) - \frac{\langle z \rangle^2}{4}  - 3f_d^2 \;,
\end{align}
which is readily inferred from the corresponding probability generating function in orders of $f_d$. Combining both results, we obtain 
\begin{align}
	\tilde{p_c}(\sigma_z) = \frac{1}{\langle z \rangle}\left(\frac{9}{4} -\frac{1}{\langle z \rangle}  + \frac{\sigma_z^2}{\langle z \rangle^2} \right)
	\label{eq:grid_exp}
\end{align} 
as a first-order approximation in $f_d$.
As the variance is quadratic in $f_d$, there are two different $f_d$, which induce the same variance but different percolation thresholds. This underlines that two moments cannot be enough to uniquely link the degree distribution to the percolation threshold. However, the first-order expansion provides an excellent approximation for small $f_d$ (see Fig.~\ref{fig:example}) and correctly captures that $p_c$ is maximized for fixed mean degree by maximizing the variance of the degree distribution, $f_d =  \frac{\langle z \rangle}{6}$. 

Thus, the percolation threshold depends on the complete distribution of vertex degrees but the first two moments already suffice for an excellent approximation. In the following, we scrutinize the circumstances under which this remains valid in a general setup.

\section{Continuum Polydispersity}
\label{sec:3}

\begin{figure}[h!]
	\centering
	\includegraphics[width=0.33 \textwidth]{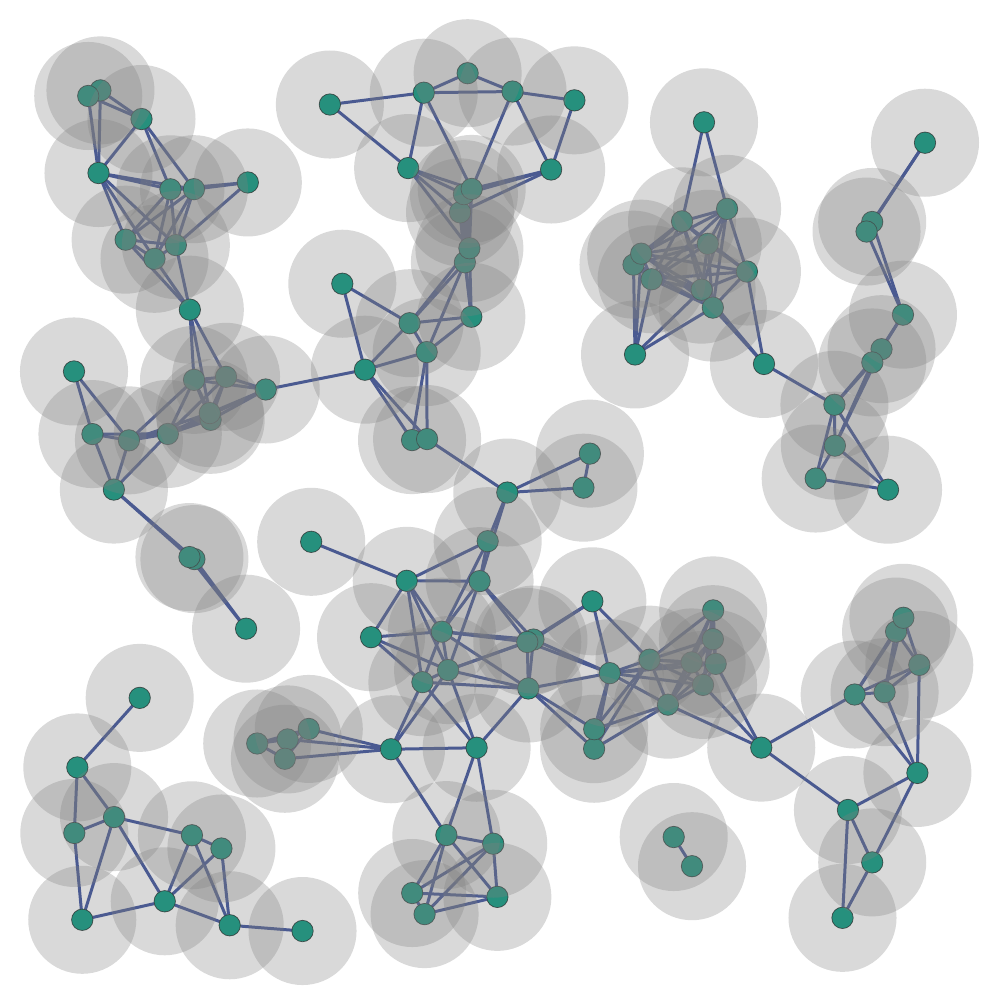}
	\caption{Exemplary network structure of a monodisperse disk system close to the percolation threshold.  }
	\label{fig:cb}
\end{figure}

In the example we were dealing with a closed equation, i.e., eq.~(\ref{eq:example}). This equation can be expanded around the monodisperse distribution, expressed in terms of the first moments of the probability distribution and solved approximately to analytically predict the relationship between polydispersity distribution and percolation threshold. 
However, in general we will not know $R$ exactly, so that we need a strategy to approximate the relevant partial derivatives. This is what we want to develop in the following for continuum percolation problems.
\newline
For the sake of concreteness, consider the simple Poisson process of non-interacting circles of diameter $d$ and number density $\rho$ in the plane that connect on intersection. Fig.~\ref{fig:cb} illustrates the system as well as the network topology emerging close to the percolation threshold. The intensity of the process is given by $\eta = \pi \rho d^2$ which means a circle has on average $\eta$ neighbors. 

Interpreting the percolation problem as a branching process we assign a circle as origin $\mathcal{O}$ and decompose the vertex set of the adjacency graph into shells characterized by the shortest distance (in terms of number of edges on the graph) to the origin.
We then define $X_n$ as the number of particles with shortest distance exactly $n$, for example $X_1$ counts the number of neighbors of the origin.
If and only if the system percolates, $R:=\lim_{n \rightarrow \infty} \expectation[X_n] >0$  \cite{coupette2022exactly}.
The infinite path length limit will generally be out of the range of what we are able to calculate analytically. 
However, we can extrapolate by treating each particle that comprises $X_n$ like the origin, neglecting correlations in the adjacency graph of range longer than $n$, i.e. loops of length longer than $2n$.  
This effectively approximates $R$ by $\lim_{k \rightarrow \infty} (\expectation[X_n])^k$, yielding

\begin{align}
	R_n:=\expectation[X_n] = 1 \; ,
	\label{eq:crit}
\end{align}
for the percolation threshold.
Importantly, as the origin can grow in all directions whereas particles on the surface of a cluster are limited in that capability, this procedure overestimates $X_n$ and thus generates a lower bound to the percolation threshold.
The approximation corresponding to $n=1$ corresponds to the so-called second virial approximation which assumes an entirely treelike topology and has been widely explored in different contexts \cite{coniglio1977pair,kyrylyuk2008continuum,meyer2015percolation}. 
It is also linked to excluded volume theory, because for a Poisson process $\expectation[X_1] = \rho V_{\mathrm{ex}}$, with $V_{\mathrm{ex}}$ being the measure of  the set of configurations of a probe particle that cause it to connect to another stationary particle, commonly referred to as excluded volume.

Now we add polydispersity to the circle system by introducing two diameters, $d_1 < d_2$, with $f$ being the relative fraction of smaller circles giving rise to partial densities of both species $\rho_1 = f \rho$ and $\rho_2 = (1-f)\rho$. We are interested in the critical density as a function of $f$ for a prescribed pair of diameters.   
Excluded volume theory now faces the issue that the excluded volume depends on the diameter of both, stationary and probe particle.
So what is the appropriate volume to describe the unique percolation threshold of the entire system?

\subsection{First order}

\begin{figure}[t]
	\centering
	\includegraphics[width=0.49 \textwidth]{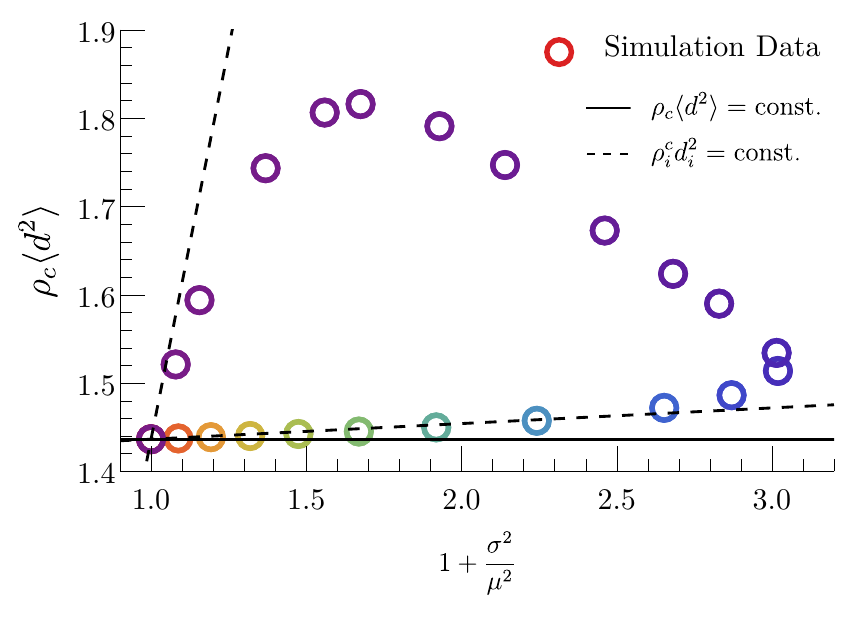}
	\caption{Dimensionless critical density of a bidisperse circle system with $d_2 = 10 d_1$ for varying relative fractions $f$ indicated by the rainbow coloring. The simulation results were taken from \cite{quintanilla2001measurement}. Red corresponds to $f=0$ and purple to $f=1$. $\mu$ and $\sigma$ denote mean and variance of the corresponding diameter distribution. The maximum critical volume fraction is reached at $f \approx 0.99$ and the relative variance is maximized at $f = \frac{10}{11}$. }
	\label{fig:hyteresis}
\end{figure}

Following the branching picture, we need to compute the average number of neighbors of a particle that is part of the cluster containing the origin. Larger circles will have more neighbors on average, hence a random circle chosen under the condition that it is another circle's neighbor is more likely to be large. 
Thus, following an arbitrary edge in the adjacency graph, we are more likely to end up at a vertex representing a large circle than a small one. This is completely analogous to the excess degree distribution experienced in the configuration model.
Concretely, the excluded volume of a particle is linear in its diameter, 
\begin{align}
	V_{\alpha \beta} = \pi \left(\frac{d_\alpha}{2}+\frac{d_\beta}{2}\right)^2 - \pi \left(\frac{d_\alpha}{2}-\frac{d_\beta}{2}\right)^2 = \pi d_\alpha d_\beta \, .
	\label{eq:Vex}
\end{align}
Notice that we discount circles that are completely contained within another circle as there is no intersection and thus not connection.
If we considered filled disks rather than circles, the corresponding configuration would lead to a connection. Naturally, the percolation thresholds of disks and circles with the same diameter distribution coincide.
As the inner circle does not provide opportunities for further growth of a cluster it does not contribute to $R$ and hence can be ignored. Thus, eq.~(\ref{eq:Vex}) also describes the relevant excluded volume for percolation of disks, though not for microscopic observables like the mean number of nearest neighbors.
The linearity of the excluded volume implies that the probability that a neighbor of any particle has diameter $d_\alpha$ is given by
\begin{align}
	\probability_\mathcal{O}(d_\alpha ) = \frac{\rho_\alpha d_\alpha}{\langle d \rangle} = \frac{\rho_\alpha d_\alpha}{\rho_1  d_1 + \rho_2 d_2} \; .
	\label{eq:PO}
\end{align} 
The subscript $\mathcal{O}$ indicates that this is the diameter distribution of the origin that we pick for the branching process, i.e., a circle that is known to have a neighbor. Generally, the probability of an arbitrary circle to have diameter $d_\beta$ is proportional to its partial density
\begin{align}
	\probability(d_\beta) = \frac{\rho_\beta}{\rho} \, .
\end{align}
The ``appropriate'' excluded volume for a bidisperse mixture of circles is therefore the weighted average
\begin{align}
	 &\sum_{\alpha,\beta} \probability_\mathcal{O}(d_\alpha) \probability(d_\beta) V_{\alpha\beta} = \sum_\beta \frac{\pi \rho_\beta d_\beta}{\rho_1+\rho_2} \sum_\alpha
	 \frac{\rho_\alpha d_\alpha^2}{\rho_1 d_1 + \rho_2 d_2}\; \nonumber \\
	 &=  \frac{\pi}{\rho_1+\rho_2} \left( \rho_1 d_1 +\rho_2 d_2 \right)\left( \frac{\rho_1 d_1^2 +\rho_2 d_2^2 }{\rho_1 d_1 +\rho_2 d_2 }\right) = \pi \langle d^2 \rangle \,.
\end{align}
We find that the excluded volume depends exclusively on the second moment of the diameter distribution. 
We can repeat the same calculation for any diameter distribution with finite second moment, discrete or continuous, with the same result. 
Recalling the criticality condition eq.~(\ref{eq:crit}) for $n=1$, corresponding to the consistent polydisperse generalization of excluded volume theory
\begin{align}
	R_1 = \rho_c V_{\mathrm{ex}} = \rho_c \pi \langle d^2 \rangle = \eta_c = 1 \;.
	\label{eq:r1}
\end{align}
Importantly, $R_1$ is generally not the mean number of neighbors of a randomly chosen particle in the system due to the diameter dependent weights of $\probability_\mathcal{O}$.
\newline
Another way we can think about this approximation is as a Markov process. The matrix
\begin{align}
	\Lambda = \pi \begin{pmatrix}
	\rho_1 d_1 d_1 & \rho_1 d_2 d_1  \\
	\rho_2d_2 d_1 & \rho_2 d_2 d_2 \\
	\end{pmatrix}
\end{align}
describes the number of neighbors of a specific size when multiplied with an initial distribution. As long as the initial distribution is not orthogonal to any of the eigenvectors of this matrix, the largest eigenvalue of $\Lambda$ describes the average reproduction rate of the process. This eigenvalue is readily computed as $\lambda_1 = \pi (\rho_1 d_1^2 + \rho_2 d2^2) = \pi \rho \langle d^2 \rangle$, leading to the aforementioned percolation threshold. In absolute terms, you will find that the prediction of this approximation is not particularly convincing as the networks formed by the circle system are far from treelike and the excluded volume is not representative for the asymptotic branching process. 
Yet, we may characterize the impact of polydispersity by expanding around the monodisperse system, which effectively means adopting the corresponding $R$-value.  However, there is no need for that as eq.~(\ref{eq:r1}) implies $\rho_c  \langle d^2 \rangle = \mathrm{const.}$. 
Therefore, within our approximation, all distributions with the same second moment (non-centralized) are equivalent which in particular includes a corresponding monodisperse system. 
Casting the second moment in terms of mean $\mu$ and variance $\sigma^2$ we obtain the continuum analogue of eq.~(\ref{eq:grid_exp}). 
Fig.~\ref{fig:hyteresis} illustrates the accuracy of the approximation. 
For a wide range of mixing ratios $f$ closer to the monodisperse system of bigger circles, the approximation is very good. Only in the range $f \in [0.9,1)$ do we seemingly miss something. 

\begin{figure*}[t!]
	\centering
	\includegraphics[width=0.49 \textwidth]{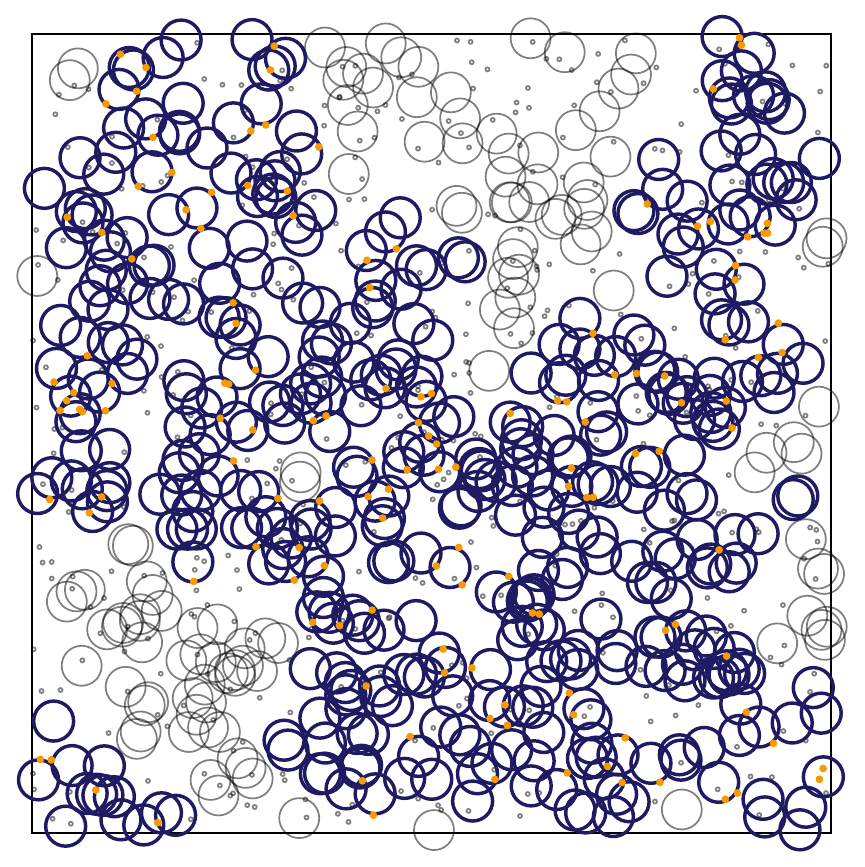}
	\includegraphics[width=0.49 \textwidth]{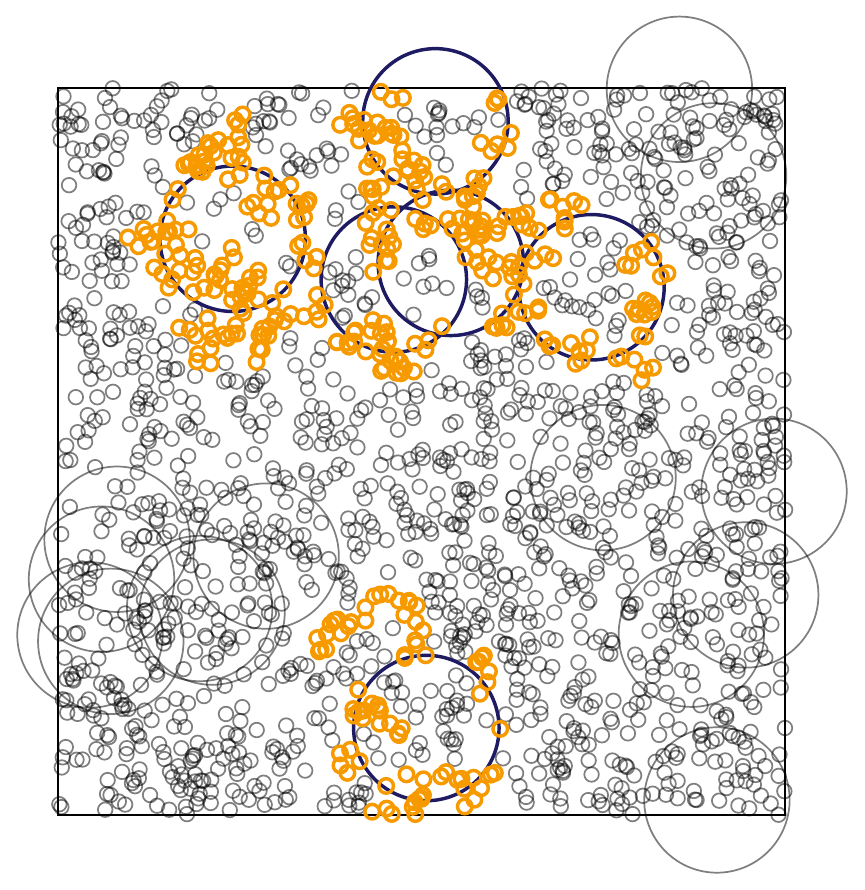}
	\includegraphics[width=0.49 \textwidth]{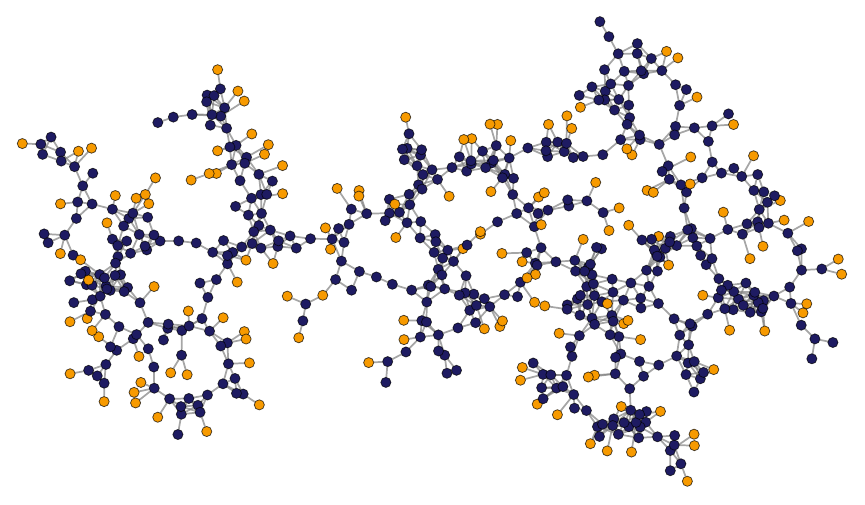}
	\includegraphics[width=0.49 \textwidth]{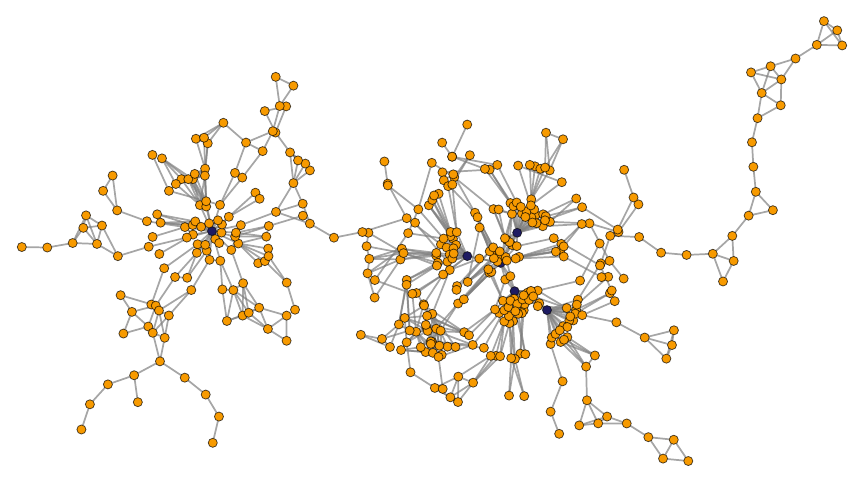}
	\caption[Topological Emulsions]{%
		Top: Topological difference for a mixture of circles with diameters $d_1$ and $10 d_1$, with fractions of smaller circles $f=0.5$ (left) and $f=0.99$ (right). 
		The largest cluster is highlighted with the color distinguishing the particle species, large (blue) and small (yellow). 
		Bottom: Connectivity graph corresponding to the snapshot above. 
		Each vertex represents a circle of matching color, and edges are drawn between intersecting circles. (Figure redrawn from \cite{coupette2023percolation})
	}
	\label{fig:top_emulsion}
\end{figure*}

The accuracy of the approximation for systems with a few small circles in a background of large ones is readily explained by the fact that the small circles do not play a role in the emergence of a giant component. Percolation essentially requires a percolating network of large circles which emerges at a density $\rho_2$ corresponding to the critical density of the monodisperse system. Accordingly, the backbone of the percolating network is very similar to that of a monodisperse system and thus adequately described with the same $R$-value. 
That changes when the large particles are so rare that islands of large particles need to be bridged by smaller ones. 
The network undergoes a topological transition akin to a phase inversion in emulsions. This transition is illustrated in Fig.~\ref{fig:top_emulsion}.
We chose the diameter ratio $\frac{d_2}{d_1} = 10$ deliberately to emphasize this transition. 
The corresponding curves for smaller ratios are entirely contained within the orbit displayed in Fig.~\ref{fig:hyteresis}. 
Depicting $\rho_c \langle d^2 \rangle$ as a function of $1+\frac{\sigma^2}{d^2}$ exaggerates the difference between the two monodisperse limits as
\begin{align}
	\frac{d \rho_c \langle d^2 \rangle}{d\left( \frac{\sigma^2}{\mu^2} \right)} = \frac{d \rho_c \langle d^2 \rangle}{df} \left( \frac{d\left( \frac{\sigma^2}{\mu^2} \right)}{df}  \right)^{-1}   \; .
	\label{eq:derivatives}
\end{align}
Since the first term vanishes, the second term does not play a role in our approximation. Yet, evaluating that expression for $f=0$ and $f=1$  yields $\frac{(d_1-d_2)^2}{d_2^2}$ and $-\frac{(d_1-d_2)^2}{d_1^2}$, respectively. Accordingly, the ratio of the slopes of the two branches emanating from the monodisperse system features a factor $\left(\frac{d_1}{d_2}\right)^2$ so that the intersection angle grows with the diameter discrepancy. 
If we slightly deviate from our branching approach and argue that a percolating cluster of large circles is indeed a necessary condition for percolation for small $f$, all the small circles become dead weight, yielding
\begin{align}
\left(	\frac{d \rho_c \langle d^2 \rangle}{df} \right)_{f=0} = \rho_{c_{|f = 0}} d_2^2  \frac{d_1^2}{d_2^2} = \frac{R_1}{\pi} \frac{d_1^2}{d_2^2} \; .
\end{align}
We can do the same thing with the roles of large and small circles reversed to obtain the analogous result for $f=1$. 
The corresponding approximations for $\rho_c \langle d^2 \rangle$ are depicted in Fig.~\ref{fig:hyteresis} as dashed lines. Clearly, the assumption corresponding to the $f=0$ branch is justified for relative fractions of up to $f \approx 0.8$. 
In the general case of a continuous distribution there is no clear distinction between small  and big particles. 
However, the observation that it takes extreme size distributions to significantly deviate from the network structure of a monodisperse system remains valid. To provide a perspective,  uniform distribution of circles within $[0,d_1]$ yields a critical value $\rho_c \langle d^2 \rangle$ within $3.5 \%$ of the monodisperse system.
It should be stressed that despite featuring the second moment of the diameter distribution, $\rho_c \langle d^2 \rangle = \mathrm{const.}$ is a first order approximation. 
As indicated by previous simulation studies, it is more appropriate to think about the first moment of the area distribution instead \cite{consiglio2004symmetry,quintanilla2007asymmetry}.  
In fact we can repeat the calculation for spheres rather than circles to find the first moment of the volume distribution to be the key quantity.
\subsection{Second order}
In order to improve our approximation we can go to the next order, i.e., construct $R_2$. That requires the computation of the average number of next nearest neighbors, which for Poisson processes is still feasible analytically. 
The property of being a next nearest neighbor (NNN) requires the existence of a particle linked to the origin $\alpha$ and the designated NNN $\beta$ simultaneously. 
The excluded volume of a disk (we have to avoid counting NNNs in the interior of the original circle) of diameter $d_\alpha$ relative to a disk of diameter $d_\gamma$ is again a disk with diameter $\frac{d_\alpha +d_\gamma}{2}$. 
Thus, the area in which a disk of diameter $d_\gamma$ intersects two other disks of diameters $d_\alpha$ and $d_\beta$, respectively, is the intersection of two disks with diameters $\frac{d_\alpha+d_\gamma}{2}$ and $\frac{d_\beta+d_\gamma}{2}$ and thus appropriately labeled ``contact lens''.
The area of this contact lens $A^{()}_{d_\alpha,d_\beta,d_\gamma}$ is a function of the center-to-center separation of $\alpha$ and $\beta$, $r = |\bs{r}_1 -\bs{r}_2|$, and depends on all three diameters. 
$A^{()}_{d_\alpha,d_\beta,d_\gamma}(r)$, though involved, is known analytically \cite{weisstein2003circle}. 
In order to avoid counting cases in which $\beta$ or $\alpha$ are completely contained in $\gamma$, we need to evaluate 
\begin{align}
	A_{\alpha\beta\gamma} &=& &A^{()}_{(d_\alpha+d_\gamma)/2,(d_\gamma+d_\beta)/2,d_\gamma} - A^{()}_{(d_\alpha+d_\gamma)/2,(d_\gamma-d_\beta)/2,d_\gamma} \nonumber \\ &-& &A^{()}_{(d_\gamma-d_\alpha)/2,(d_\gamma+d_\beta)/2,d_\gamma} +
	A^{()}_{(d_\gamma-d_\alpha)/2,(d_\gamma-d_\beta)/2,d_\gamma} \,.
\end{align}
All these areas are functions of the same separation $r$.
As there are no correlations between particle positions, the probability that an area is entirely devoid of particles of density $\rho_\gamma$ is $\exp(- \rho_\gamma A_{\alpha \beta \gamma})$. 
Accordingly, the probability that a particle at distance $r$ from $\alpha$ is indeed a next-nearest neighbor is given by
\begin{align}
\probability_{\alpha \beta}:=\probability(\beta \text{ is NNN of } \alpha) = 1 - \exp\left(- \sum_\gamma \rho_\gamma A_{\alpha \beta \gamma}\right)
\end{align}
We integrate this probability over all possible locations for a diameter $d_\beta$ that do not cause a direct intersection with $\alpha$ yielding the average number of NNNs with diameter $d_\beta$ of a disk of diameter $d_\alpha$
\begin{align}
	R_{\alpha \beta} = 2 \pi  \rho_\beta \int^{\infty}_{\frac{d_\alpha + d_\beta}{2}} \dd r \, r  \, \probability_{\alpha \beta} \; .
\end{align}
Ultimately, we can again construct a Markov model with transition matrix
\begin{align}
	\Lambda =  \begin{pmatrix}
		R_{11} & R_{21}  \\
		R_{12} & R_{22} \\ 
	\end{pmatrix} \; ,
\end{align}
and compute the largest eigenvalue which yields $R_2$. 
\newline
\begin{figure}[t]
	\centering
	\includegraphics[width=0.49 \textwidth]{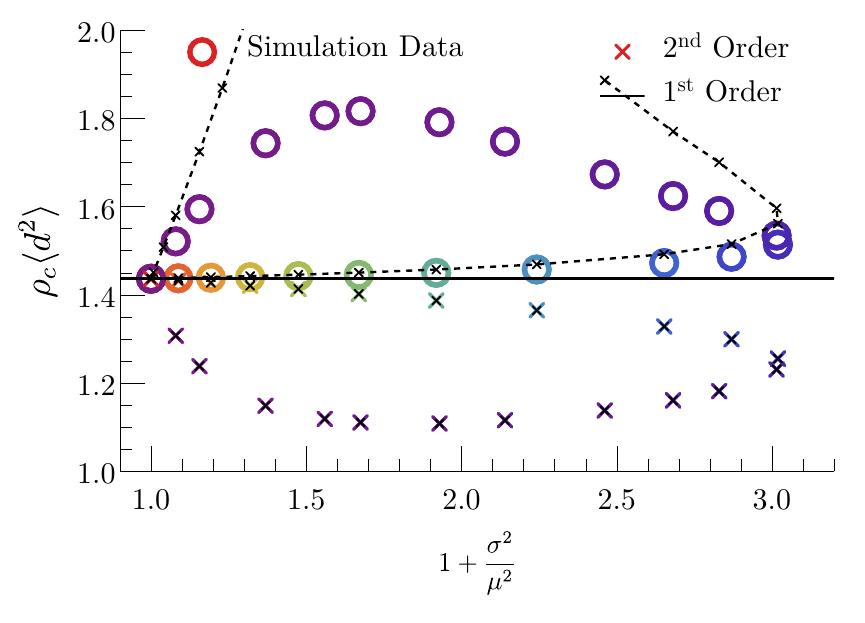}
	\caption{As Fig.~\ref{fig:second_order}, with the $\mathrm{2^{nd}}$ order predictions included (colored crosses). Color again indicates the mixing ratio $f$ running from red ($f=0$) to purple ($f=1$). The dashed lines correspond to the second order prediction as a result of demanding a percolating cluster of large circles (bottom branch) and small circles (top branch), respectively.  }
	\label{fig:second_order}
\end{figure} 
\newline
By construction \cite{coupette2023universal}, the absolute value of the prediction corresponding to $R_2 = 1$ has become more precise and remains a rigorous lower bound. While the first order approximation predicts a critical value $\rho^1_c d^2 = {\pi^{-1}} \approx 0.318$ for the monodisperse system compared to $\rho_c d^2 \approx 1.436$ in simulations, the second order provides a ``significant'' improvement to $\rho_c d^2 \approx 0.532$. Naturally, after including loops of size four in the adjacency graph, we cannot expect too much in view of the actual topology of the monodisperse network (cf. \ref{fig:cb}). Nevertheless, following the protocol introduced in  \cite{coupette2023universal}, i.e.~restricting the branching process to a half space and associating criticality with $R_1 = R_2$, we observe an actual improvement to $\rho_c d^2 \approx 1.097$.
Moreover, we obtain a non-trivial dependence on the diameter distribution as we expand around the monodisperse system and we can redraw Fig.~\ref{fig:hyteresis} with our second order predictions (cf. Fig.~\ref{fig:second_order}).
We obtain a very similar curve as found in the simulations but unfortunately flipped with respect to the result of the monodisperse system. Instead of a general increase in the critical area fraction we find a decrease with a very similar absolute difference to the monodisperse system but flipped sign. As a consequence, we predict a minimal area fraction where the simulations show a maximum. How can this be?
\newline
\newline
The lower predictions for the percolation threshold are due to $R_2$ increasing with polydispersity rather than decreasing. We observe that the average number of next nearest neighbors across all the bidisperse distributions is actually minimal for the monodisperse system. Taking a second look at the adjacency graph for $(f=0.5)$ in Fig.~\ref{fig:top_emulsion} we can see why -- the largest cluster is an almost percolating network of large circles with the majority of small circles sitting on the ``outside'' of the cluster acting as dead ends, i.e. not providing additional interconnections for the network. Yet, all these small circles inflate the statistics of next nearest neighbors. In a higher order calculation, we would eventually discard all dead ends. Indeed, for $f=0.5$, the sample snapshot suggests that the third order would already be enough. If we branch from a large circle $\alpha$ to a small one $\beta$, a significant part of the small circle's future prospects of branching is denied. With respect to other small circles $\gamma$, $\beta$ simply has a very small cross section. However, if $\gamma$ is large, configurations that feature an immediate connection between $\alpha$ and $\gamma$ are likewise excluded as they would have been integrated out in the same step that connected $\beta$. We neglect all correlations between a nearest neighbor and its branching history which in second order includes the origin as well as its neighbors. In that sense we double down on our neglect when going to the second order. That said, if we go to sufficiently high order, the system will eventually become treelike in the subcritical vicinity of the percolation threshold rendering a branching analysis arbitrarily precise.  Generally, if there is interest in higher order estimates (through simulation for example) it is advisable to integrate out portions of space rather than neighborhoods in an adjacency graph as this sets a palpable euclidean length scale for the neglected correlations. 
\newline
\newline
In practical application, even the second order will often stretch the boundaries of the analytically feasible. Yet, the second order still grants an important insight. The critical $R_1 = \rho_c \pi \langle d^2 \rangle$ is approximately constant around the percolation threshold which means the critical number of connections in the system varies only slowly with the diameter distribution. Close to $f=0$, the addition of small particles induces local bunching as the number of large next-nearest neighbors drops in favor of more small next-nearest neighbors which by design are on average much closer to the origin of the branching process. Thus, we ``spend'' more connections locally which inhibits percolation. Similarly, close to $f=1$ the addition of a few big particles induces hubs in the network structure (cf. \ref{fig:top_emulsion}). This increases the number of nearest neighbors but leads to a strongly inhomogeneous degree distribution which again obstructs percolation. 
Accordingly, $R_2$ measures the local ineffectiveness of the network leading to the curious anti-correlation depicted in Fig.~\ref{fig:second_order}. 

However, a second order calculation often will not be worth the effort because the calculations quickly become involved, are highly system specific, and frequently only yield marginal improvements. Nevertheless, it is good to have a framework to systematically analyze the nuanced imprint of particle polydispersity, even if often impractical. 
In contrast to that, the first order approximation is beautifully simple and, as will delineate in the remainder of this paper, surprisingly universal.

\section{The practical part}
\label{sec:4}

So far, we focused on very specific systems. There is a good reason for the circle system we picked to demonstrate our considerations: it exhibits behavior that visibly exceeds our first order approximation. Yet, bear in mind that even in that model we had to go to extremely asymmetric distributions as $\rho_c  \langle d^2 \rangle = \mathrm{const.}$ remains an excellent approximation for the rest. It is a good approximation because small particles play a negligible role in the network formation. And the role of small particles becomes even smaller as we increase the dimension of space as the excluded volume scales with a higher power of the particle size. Likewise, shape is essentially irrelevant as the scaling of the excluded volume with the corresponding size quantifier does not change with shape. Finally, even particle correlations induced pair pair-interactions have hardly an impact as long as the polydispersity does not trigger a thermodynamic phase transition. But one thing at a time.

\subsection{Shape}

\begin{figure}[b]
	\centering
	\includegraphics[width=0.49 \textwidth]{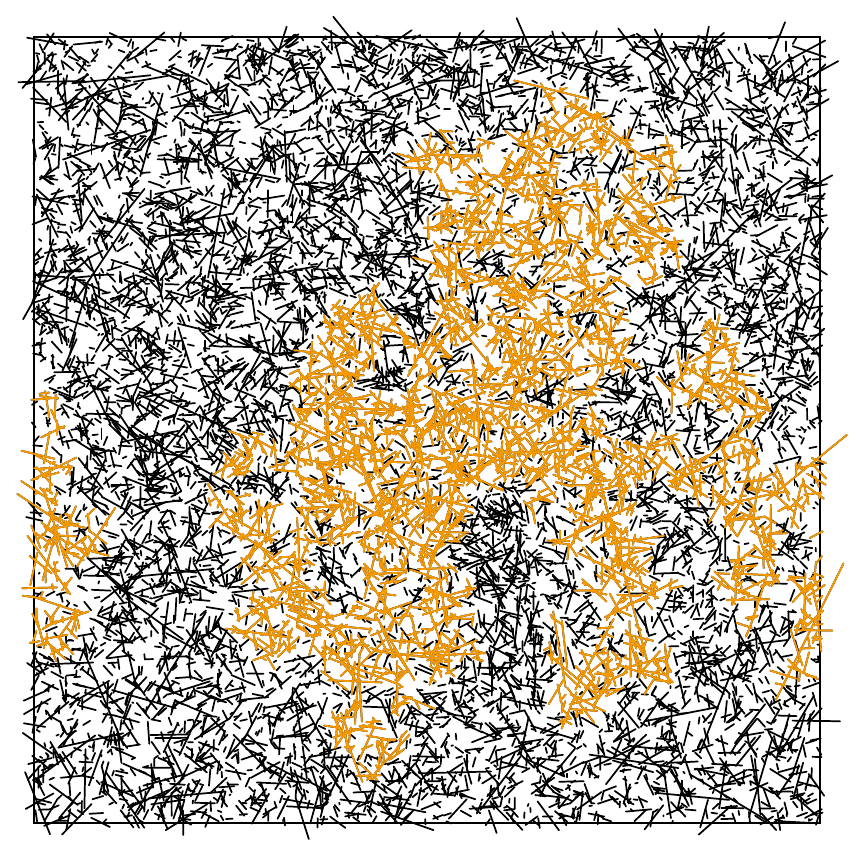}
	\caption{Snapshot of a line ensemble drawn from a log-normal distribution with the largest cluster highlighted. Figure redrawn from \cite{coupette2023percolation} .}
	\label{fig:lof_sticks}
\end{figure}
\begin{figure}[t]
	\centering
	\includegraphics[width=0.49 \textwidth]{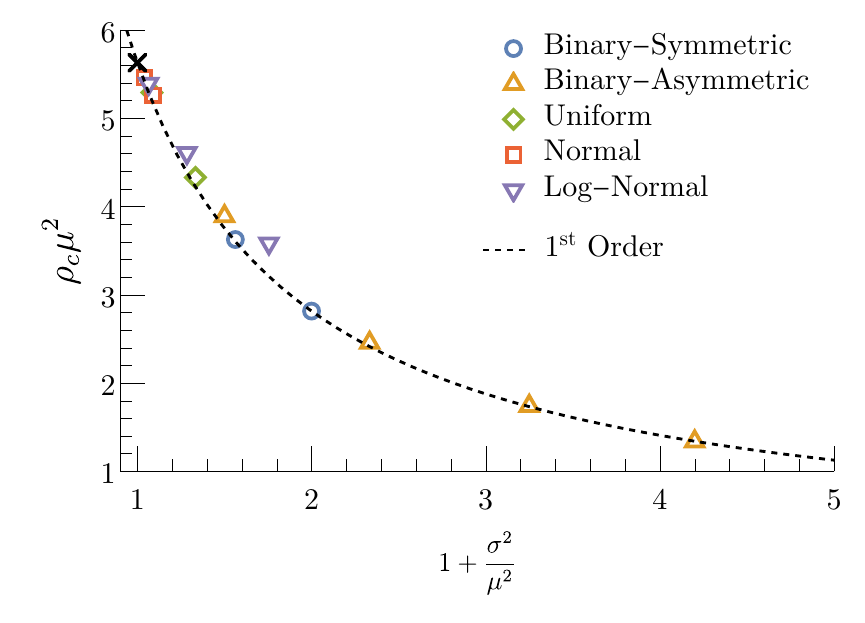}
	\caption{Critical density for a variety of polydisperse line ensembles as indicated in comparison to the first order approximation $\rho \langle l^2 \rangle = \mathrm{const.}$. Notice that we scaled the density with the mean of the distribution $\mu$ rather than the second moment providing a different perspective. The cross marks the monodisperse system. }
	\label{fig:polyrod}
\end{figure}

 
In terms of aspect ratio, a line segment seems to be the opposite of a circle. A line segment is described by its length $l$, its center position, as well as its orientations which we may parameterize with the angle $\varphi$ relative to the x-axis. We treat lines as connected if they intersect. Then, the excluded area of one line segment of length $l_\alpha$ with respect to another with length $l_\beta$ at an relative angle $\phi$ is a rhombus with area
\begin{align}
	A_{\mathrm{rh}}(\varphi) = l_\alpha l_\beta \sin(\varphi) \; . 
\end{align}
Given an isotropic distribution of orientations we average over all orientations yielding the mean excluded volume of two line segments \cite{bug1986continuum}
\begin{align}
	V_{\alpha \beta} = \frac{2}{\pi} l_\alpha l_\beta \; .
\end{align}
This simply is the same expression as for the circles only with a different prefactor.
Hence all the considerations discussed in the contect of circles can be transferred to line segments without any further thought. We expect that all length distributions with the same second moment are in good approximation equivalent and Fig.~\ref{fig:polyrod} confirms that expectation.

The only systems to substantially deviate from our prediction are the log-normal distributions. Those distributions feature rare instances of lines that exceed the average size by orders of magnitude (cf. Fig.~\ref{fig:lof_sticks}). Those rods naturally interconnect an entire sub-domain and become hubs of the emerging network, skewing the degree distribution akin to the $f \rightarrow 1$ limit of the bidisperse circle system. Thus, we understand the mechanism driving the deviation but a thorough prediction for the family of log-normal distributions would require a higher order analysis. In view of a cost-benefit analysis we refer to heuristic fit functions \cite{tarasevich2018percolation}. Notice, that in absolute terms, the first order prediction corresponding to $R_1 = 1$ is even worse than for circles as the network formed by lines deviates from a tree even more thoroughly than the circle system. To obtain an quantitatively accurate prediction for this model we need to resort to different means \cite{coupette2021nearest}. 

\subsection{Dimension}

\begin{figure}[t]
 	\centering
 	\includegraphics[width=0.49 \textwidth]{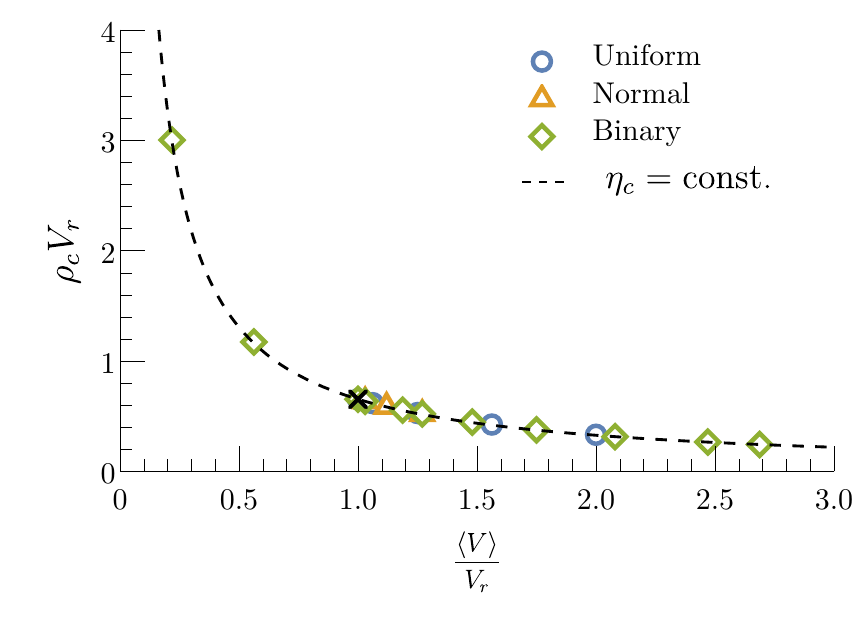}
 	\caption{Critical density for a variety of polydisperse sphere ensembles as indicated alongside our first order approximation $\rho \langle l^2 \rangle = \mathrm{const.}$. $V_r$ is the reference volume of a monodisperese system with unit diameter. The cross marks the monodisperse system. }
 	\label{fig:polysphere}
 \end{figure}

Percolation problems tend to become simpler with increasing spatial dimension. This does not only apply for the critical scaling behavior which is of mean field type above the upper critical dimension but also for the percolation threshold itself \cite{grimaldi2015continuum}. The prime example for that is the three-dimensional analogue of the stick system that we analyzed in the previous section. Slender rods form approximately treelike networks which renders the percolation problem in the so-called Onsager limit exactly solvable. In this limit, the first order approximation, in this context usually referred to as second virial approximation, becomes even quantitatively accurate \cite{kyrylyuk2008continuum,otten2009continuum,otten2011connectivity}. This naturally implies that polydispersity can well be treated by the same approximation which previous studies have already explored.
Spheres are hence the more challenging end of the anisotropy spectrum. Yet, conceptually this system is very similar to the circles, the only change being the scaling of the excluded volume
\begin{align}
	V_{\alpha \beta} = \frac{\pi}{6} \left((d_\alpha + d_\beta)^3 - |d_\alpha - d_\beta|^3  \right) \; .
	\label{eq:sphereV}
\end{align} 
With that we can determine the largest eigenvalue $\lambda_1$ of the corresponding Markov matrix $\Lambda$. Due to the cross terms, the average excluded volume is in general not exactly the first moment of the volume distribution, i.e., the third moment of the diameter distribution. However, the relative deviation for the distributions we are interested in is negligibly small. Thus, we suspect that all distributions with the same mean volume $\frac{\pi}{6} \langle d^3 \rangle$ are equivalent which means the critical volume fraction $\eta = \rho_c \frac{\pi}{6} \langle d^3 \rangle$ is conserved. As Fig.~\ref{fig:polysphere} shows, the first order is once more spot on.
Since the cubic scaling amplifies the relative dominance of the largest structures, challenging the approximation requires even more thoroughly skewed distributions compared to the circle system. Standard continuous choices such as normal or uniform distribution are readily mapped onto the monodisperse system. There is a  surprising shortage of simulation data for this simple system, presumably because it is requires a lot of effort to find even the small deviations from the simplicity of a conserved critical volume fraction. In \cite{consiglio2004symmetry} a strongly skewed binary distribution maximized the relative deviation among all distributions essayed  -- the maximal deviation was found to be roughly $3.2 \%$. 

\subsection{Hard Interaction}

\begin{figure}[t]
 	\centering
 	\includegraphics[width=0.49 \textwidth]{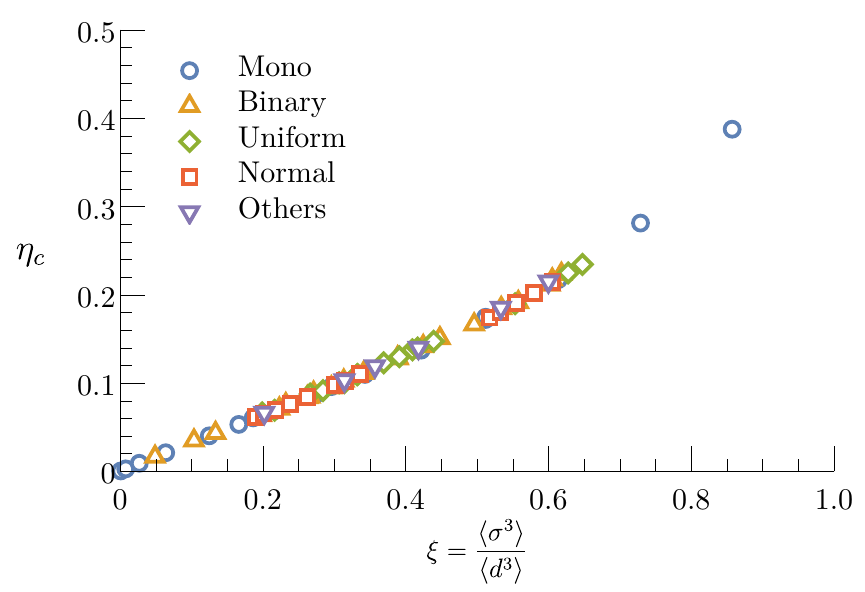}
 	\caption{Critical hard volume fractions for a variety of polydisperse hard sphere ensembles sampled via standard Monte Carlo simulation. Within a given ensemble we fixed the thickness of the connectivity shell $d-\sigma$ for each individual particle mimicking a characteristic tunneling length.
 		``Others'' include log-normal distributions as well as discrete distributions with more than two particle species. }
 	\label{fig:polyhs}
 \end{figure}

In a final attempt to break the $\mathrm{1^{st}}$ order approximation for spheres, we introduce a hard repulsion between a hard core of diameter $\sigma$ accompanied by a larger concentric connectivity shell $d$ ("cherry pit model"). Both, $\sigma$ and $d$ can be subject to polydispersity. As before, we need to evaluate the average number of neighbors of a sphere of a given diameter to construct the matrix $\Lambda$. The excluded volume is a straightforward generalization of \ref{eq:sphereV}
\begin{align}
	V_{\alpha \beta} = \frac{\pi}{6} \left((d_\alpha + d_\beta)^3 - (\sigma_\alpha + \sigma_\beta)^3  \right) \; .
	\label{eq:hsphereV}
\end{align} 
However, due to the hard core repulsion we have thermodynamic correlations in the system captured by the species-dependent pair-distribution function $g_{\alpha \beta}$. We obtain the average number of nearest neighbors for a given pair of hard spheres by integrating $\rho_\beta g_{\alpha \beta}$ over the excluded volume. Hence, the interaction induces a density dependent weight function to the calculation. This weight could be determined through sophisticated means like fundamental measure theory. Yet, for lower densities we expect $g _{\alpha\beta}= \Theta(r-\frac{1}{2}|\sigma_\alpha +\sigma_\beta|)$ to be a decent approximation, so that we immediately return to the plain excluded volume. Just like for the ideal spheres, the mixing terms cause formal complications but $\langle d^3 \rangle + \langle \sigma^3 \rangle$ is a very good approximation of the largest eigenvalue of the corresponding Markov matrix. Thus, we expect the ratio $\xi = \frac{ \langle \sigma^3 \rangle}{ \langle d^3 \rangle}$ to be the characteristic quantity that causes the hard critical volume fraction $\eta_c = \frac{\pi}{6}\rho_c \langle \sigma^3 \rangle$ different hard core and connectivity shell size distributions to collapse onto a common curve. Despite our crude treatment of the thermodynamics, Fig.~\ref{fig:polyhs} convincingly confirms our hypothesis.
In the vicinity of a phase transition such a the freezing transition in the hard sphere system we expect our approach to break down as the entire mapping on the branching process relies on macroscopic randomness. Yet, for reasonably dilute systems percolation is very much indifferent towards the specific shape of the size distribution. This even applies for much more complicated particle shapes as our recent simulation study of rigid fractal aggregates underlines \cite{coupette2021percolation}. From an engineering  point of view, this forms the key observation as it suggests that the polydispersity that is present and unavoidable in most realistic scenarios ultimately has a very controlled impact. In fact, we may even think of inverting the problem using percolation to characterize polydispersity.

\section{Conclusion}

Percolation is about the behavior of the big particles. We found that for most size distributions polydispersity does not substantially alter the network structure as small particles play a negligible role in the formation of the percolating cluster. As a consequence, the percolation threshold can be inferred from a monodisperse reference system. This implies that the critical density scales as the inverse of the microscopic excluded area or volume averaged over the corresponding size distribution. This is the common denominator of the well established behavior of polydisperse ensembles of rodlike particles, fractal aggregates, hard spheres, circles, line segments, and planar lattices. We have demonstrated how to construct the first order approximation that links size distribution and critical parameters and goes as second virial, excluded volume or tree approximation, respectively, depending on context. Moreover, we outlined how to systematically improve on this approximation for the rare case of substantially skewed distributions for which the smaller structures are significant.

\begin{acknowledgments}
	\noindent
We acknowledge funding by the German Research Foundation in the project 457534544. 
\end{acknowledgments}

\section*{Author Contributions}
\noindent
FC: conceptualization (equal), investigation \& writing \\
TS: conceptualization (equal) \& funding acquisition

\section*{Data Availability}
\noindent
All simulation results are available on reasonable request to Fabian Coupette.

\section*{Conflict of Interest}
\noindent
There is no conflict of interest to declare.

\end{document}